\documentclass[preprint]{aastex}
\begin{document}
\def\etal{{\it et al.\/}}
\def\cf{{\it cf.\/}}
\def\ie{{\it i.e.\/}}
\def\eg{{\it e.g.\/}}

\title{On the universality of the spectrum of cosmic rays accelerated
at highly relativistic shocks}
\author{{\bf Mario Vietri}}
\affil{
Universit\`a di Roma 3, Via della Vasca Navale 84, 00147 Roma, \\
Italy, E-mail: vietri@amaldi.fis.uniroma3.it \\
}

\begin{abstract}
I consider analytically particle acceleration at relativistic shocks, in the 
limit of pitch angle diffusion and large shock Lorentz factors.
The derived energy spectral index ($k = (1+\sqrt{13})/2
\approx 2.30$), and particle pitch angle distribution at the shock are 
successfully compared with the results of numerical solutions. This totally 
analytic derivation is completely independent of the detailed dependence
of the diffusion coefficient $D(\mu,p)$ on either parameter, and it is
argued on a physical basis that the orientation of the magnetic field is also
irrelevant, making $k$ a universal index.
\end{abstract}


\section{Introduction}

There is currently a growing interest in the acceleration of non--thermal
particles at highly relativistic shocks. There are three classes of 
{\it bona fide} relativistic sources: beyond the well--established 
extra--Galactic (Blazars) and Galactic (superluminal) sources, both of which 
exhibit superluminal motions, 
it is now also well-established that Gamma Ray Bursts (GRBs) display
highly relativistic expansions, with Lorentz factors well in excess of 
$100$. Other classes of relativistic sources may include 
Soft Gamma Ray Repeaters (SGRs), whose recurrent explosions are largely 
super--Eddington, and special SNe similar to SN 1998bw, which displayed
marginally Newtonian expansion ($\approx 6\times 10^4\; km\; s^{-1}$) when 
optical emission lines became detectable, about a month after the explosion. 

With the discovery of GRBs' afterglows, it has now become feasible to 
derive the energy spectral index $k$ of electrons accelerated at the forward
shock, as a function of the varying (decreasing) shock Lorentz factor 
$\gamma$, provided simultaneous wide--band spectral coverage is available. With 
the launch of the USA/Italy/UK mission SWIFT, these data will become  available 
for a statistically significant number of bursts, testing directly models for
particle acceleration at relativistic shocks. Furthermore, since GRBs must also
clearly accelerate protons, the same index $k$ may determine the spectrum
of ultra high energy cosmic rays observed at Earth. 

However, until recently, both the lack of astrophysical motivation and the 
difficulty inherent in treating highly anisotropic distribution functions have 
stiffened research on this topic. Early work, both semi--analytic
and outright numerical, has concentrated on barely relativistic flows with
Lorentz factors of a few, well suited to Blazars and Galactic superluminals,
but clearly insufficent for GRBs, the only exception being the numerical
simulations of Bednarz and Ostrowski (1998). It is the purpose of this {\it 
Letter} to perform an analytic investigation of the large $\gamma$ limit, to 
establish which (if any) of the properties of the particles' distribution 
function depend upon the physical conditions of the fluid. 

\section{The analysis}

I deal first with pure pitch angle scattering, and then (Subsection 2.3) I
discuss oblique shocks. In the well--known equation for the particles' 
distribution function, under the assumption of pure pitch angle scattering,
\begin{equation}
\label{main}
\gamma(u+\mu)\frac{\partial f}{\partial z} = \frac{\partial}{\partial
\mu} \left(D(\mu, p) (1-\mu^2)\frac{\partial f}{\partial \mu}\right) \;,
\end{equation}
$f$ is computed in the shock frame, in which are also defined the 
distance from the shock $z$, the fluid speed in units of $c$, $u$, and 
fluid Lorentz factor $\gamma$. Instead, the scattering coefficient $D$, 
particle momentum $p$ and particle pitch angle cosine, $\mu$, are all defined
in the local fluid frame. I make no hypothesis whatsoever about $D$, except
that it is positive definite and smooth. We place ourselves in the shock frame,
and call $z =0$ the shock position; the upstream section is for $z < 0$, so that
the fluid speeds are both $> 0$. The above equation admits of an integral: by 
integrating over $\mu$ and $z$ we see that
\begin{equation}
\label{conserved1}
\int_{-1}^{1} (u+\mu) f d\!\mu = \mbox{const.}\;,
\end{equation}
independent of $z$. The required boundary condition for $f$, \ie, that 
$f \rightarrow 0$ as $z \rightarrow -\infty$, implies that the constant, 
upstream, is $0$. Downstream, Eq. \ref{conserved1} is also a constant, but, 
because of Taub's jump conditions, it is not the same constant as upstream. 
Since the required boundary condition for $f$ far downstream ($f_\infty$)
is that it becomes isotropic and $f$ is a relativistic invariant, we see that, 
far downstream, $\int_{-1}^{1} \mu f_\infty d\!\mu = 0$. We thus have 
\begin{equation}
\label{conserved2}
\int_{-1}^{1} (u+\mu) f d\!\mu = \left\{
\begin{array}{ll}
0 & z < 0 \\
\int_{-1}^{1} u f_\infty d\!\mu > 0 & z > 0
\end{array}
\right.
\end{equation}
where the inequality in Eq. \ref{conserved2}b (which will become necessary 
later on) derives from the obvious constraint $f > 0$. 

\subsection{Upstream}

I begin the analysis by considering Eq. \ref{conserved2}a in the limit of
very large shock Lorentz factors, in which case, upstream, $u \rightarrow 1$.
For $u=1$, this reduces to 
\begin{equation}
\int_{-1}^1 (1+\mu) f d\!\mu = 0 \;.
\end{equation}
Since $1+\mu > 0$ everywhere in the integration interval except of course at
$\mu = -1$, and since $f\geq 0$ we see that, 
for $u = 1$ we must have $f \propto \delta(\mu+1)$ where $\delta(x)$ is 
Dirac's delta. Thus, in this limit, the angular dependence factors out. 
For reasons to be explained in the next Subsection, we shall also need $f$
for $ 1-u \ll 1$, but still $\neq 0$. To search for such a solution, we let
ourselves be guided by the solution at $u =1$: thus, we let the angular
dependence factor out, and use the {\it Ansatz} $f = g(z) w((\mu+1)/h(u))$. 
Here $h(u)$ is an as yet undetermined function of the pre--shock fluid speed
such that $h(u) \rightarrow 0$ as $u \rightarrow 1$. In this way, as the speed 
grows, the angular dependence becomes more and more concentrated toward 
$\mu = -1$, as required by the previously found solution for $u = 1$. 
Introducing our {\it Ansatz} into Eq. \ref{main} I find
\begin{equation}
\frac{\gamma}{g} \frac{d g}{d z} = 
\frac{1}{(1+\mu)w}  
\frac{d}{d\mu} \left( D(\mu,p)(1-\mu^2)\frac{d w}{d \mu}\right)
= \frac{2 D_{-1} \lambda^2}{h^2(u)}\;,
\end{equation}
where I defined $D_{-1} \equiv D(\mu = -1, p)$, and I factored the eigenvalue
for future convenience. Concentrating on the angular part, and defining 
$(\mu+1)/h(u) \equiv y$, and $\dot{w} \equiv dw/dy$, $\ddot{w} \equiv 
d^2 w/d y^2$, I find
\begin{equation}
\label{down}
\frac{D(\mu,p)(1-\mu^2)}{h^2(u)} \ddot{w} + 
\frac{\dot{w}}{h(u)}\frac{d}{d\mu} \left(D(\mu,p)(1-\mu^2)\right) - 
\frac{2\lambda^2 D_{-1}(1+\mu) w}{h^2(u)} = 0\;.
\end{equation}
We are interested in a solution of the above equation only in the limit
$h(u)\rightarrow 0$, the only one in which our factored {\it Ansatz} is a good 
approximation to the true $f$. In this case, the term $\propto \dot{w}$ is 
clearly subdominant, and can be neglected in a first approximation (this 
technique is called dominant balance, Bender and Orszag 1978). Furthermore,
for $h(u)\rightarrow 0$, we expect $w(\mu)$ to be more and more concentrated 
around $\mu = -1$, so that in this range we can approximate the term 
$D(\mu,p)(1-\mu^2) \approx 2 D_{-1} (1+\mu)$, and I obtain $\ddot{w} \approx 
\lambda^2 w$ with obvious solution $w \approx w_0 \exp{(-\lambda 
(\mu+1)/h(u))}$. The factor $\lambda/h(u)$
can be determined by inserting this approximate expression for $w$ into Eq. 
\ref{conserved2}a. A trivial computation yields $\lambda/h(u) = 1/(1-u)$. 

Now, going back to the equation for $g$, the spatial part of $f$, we find,
also using the above,
\begin{equation}
\frac{1}{g} \frac{d g}{d z} \approx  \frac{2 D_{-1}}{\gamma (1-u)^2}
\approx 8\gamma^3 D_{-1}
\end{equation} 
from which, in the end, I find an approximate solution for the distribution
function in the limit $u\rightarrow 1$:
\begin{equation}
\label{up}
f \approx A \exp{(8 \gamma^3 D_{-1} z}) \exp{\left(- (\mu+1)/(1-u)\right)}\;.
\end{equation}
It is thus seen that the detailed shape of the pitch angle scattering 
function $D(\mu, p)$ is irrelevant, and that what is left of it (its value
$D_{-1}$ at $\mu = -1$) only enters the spatial part of the distribution
function $f$, not the angular one. 

\subsection{Downstream}

We make here the usual assumption, that the distribution function depends
upon the particle momentum $p$ as $f \propto p^{-s}$ in either frame 
(but see the Discussion for further comments). From the condition
of continuity of the distribution function at the shock, denoting as $p_a$ and 
$\mu_a$ the particle's momentum and cosine of the pitch angle in the downstream 
frame, we have
\begin{equation}
\frac{1}{p_a^{s}} w_a(\mu_a) \propto 
\frac{1}{p^{s}} \exp{(-(\mu+1)/(1-u))}
\end{equation}
where the irrelevant constant of proportionality does not depend on $p, p_a,
\mu, \mu_a$. Using the Lorentz transformations to relate $p, p_a, \mu, \mu_a$
($\mu = (\mu_a - u_r)/(1-u_r\mu_a)$, $p = p_a \gamma_r (1-u_r \mu_a)$, 
with $u_r$ and $\gamma_r$ the relative speed and corresponding Lorentz factor
between the upstream and downstream fluids), I find
\begin{equation}
w_a(\mu_a) = \frac{1}{(1-u_r \mu_a)^s} \exp{\left(-
\frac{(\mu_a+1)(1-u_r)}{(1-u)(1-u_r \mu_a)}\right)}\;.
\end{equation}
For $u \rightarrow 1$, it is easy to derive from Taub's conditions (Landau
and Lifshitz 1987) that $u_r\rightarrow 1$, and that $(1-u_r)/(1-u) \approx 
\gamma^2/\gamma_r^2 \rightarrow 2$. This result does use a post--shock 
equation of state $p = \rho/3$, which is surely correct in the limit $u
\rightarrow 1$. In the end, I obtain
\begin{equation}
\label{incomplete}
w_a(\mu_a) = \frac{1}{(1-\mu_a)^s} \exp{\left(-2
\frac{\mu_a+1}{1-\mu_a}\right)}\;.
\end{equation}
This equation shows why we needed to determine the pitch angle distribution,
in the upstream frame, even for $1-u \neq 0$: in fact, even though the
angular distribution in the upstream frame (Eq. \ref{up}) tends to a 
singularity, the downstream distribution does not (because the factor 
$(1-u)/(1-u_r)$ has a finite, non--zero limit), and the concrete form to 
which it tends depends upon the departures of the upstream distribution from a
Dirac's delta. 

From now on I will drop the subscript $a$ in $\mu_a$, since all quantities
refer to downstream.
In order to determine $s$, we now appeal to a necessary regularity condition 
which must be obeyed by the initial (\ie, for $z = 0$) pitch angle distribution,
Eq. \ref{incomplete}. Looking at Eq. \ref{main} specialized to the downstream 
case, where $u = 1/3$ for very fast shocks, we see that this equation has a 
singularity at $\mu = -1/3$. Passing through this singularity will fix the 
index $s$. 
It is not convenient to use $f$ directly; rather, I use its Laplace transform
\begin{equation}
\label{laplace}
\hat{f}(r,\mu) \equiv \int_0^{+\infty} f(z,\mu) e^{-rz} d\!z\;.
\end{equation}
Taking Laplace transforms of both sides of Eq. \ref{main} I obtain
\begin{equation}
\label{transformed2}
-\frac{\gamma (1/3+\mu)w_a(\mu)}{r} + \gamma (1/3+\mu) \hat{f} = 
\frac{1}{r} \frac{\partial }{\partial 
\mu} \left(D(\mu,p)(1-\mu^2) \frac{\partial\hat{f}}{\partial\mu}\right)\;.
\end{equation}

I am interested in the limit $r \rightarrow +\infty$. In fact, here I can use
two results. First, in this limit, it is well--known (Watson's Lemma, Bender
and Orszag 1978) that Eq. \ref{laplace} reduces to 
\begin{equation}
\label{watson}
\hat{f}(r,\mu) \rightarrow \frac{f(z=0,\mu)}{r} = \frac{w_a(\mu)}{r}\;.
\end{equation}
Despite this wonderful result in all its generality, I will actually use it
only in the neighborhood of $\mu = -1/3$; here, Eq. \ref{transformed2} takes 
on a simple form: defining $t \equiv \mu + 1/3$, 
\begin{equation}
\label{localized}
\frac{b t}{r} + a t \hat{f} = \frac{1}{r} \frac{\partial^2 \hat{f}}{\partial 
t^2} + \frac{c}{r}\frac{\partial \hat{f}}{\partial t}
\end{equation}
where I defined $b \equiv (\gamma w_a(\mu) D(\mu,p)(1-\mu^2))|_{\mu =-1/3}$, 
$a \equiv (\gamma D(\mu,p)(1-\mu^2))|_{\mu =-1/3}$, and $c \equiv 
(\partial/\partial\mu D(\mu,p)(1-\mu^2))/D(\mu,p)(1-\mu^2)|_{\mu = -1/3}$. 
Now I make the {\it Ansatz} (to be checked {\it a posteriori}) that the term 
$\propto\partial\hat{f}/\partial t$ is negligible compared to the second order
derivative in the limit $r\rightarrow +\infty$. I am interested only in the
most significant term in $r$, since Eq. \ref{watson} was only obtained to 
this order. Then, the equation \ref{localized} becomes
\begin{equation}
\label{model}
a t \hat{f} = \frac{1}{r}\frac{\partial^2 \hat{f}}{\partial t^2}\;.
\end{equation}
The above equation is the prototype of the 
one--turning point problem. Its solution, strictly in the neighborhood of the 
point $t=\mu+1/3=0$, is (Bender and Orszag 1978, Sect. 10.4, Eq. 10.4.13b):
\begin{equation}
\label{airy}
\hat{f}(t) \approx r^{1/12} C Ai(r^{1/3} a t)
\end{equation}
where $C$ is an arbitrary constant, and $Ai(x)$ is one of Airy's functions. 
From this it can easily be checked that our {\it Ansatz} was justified. 

Clearly, close to the point $t=\mu+1/3=0$, Eq. \ref{watson} and Eq.
\ref{airy} must give the same results. Thus I find that, close to $\mu = -1/3$,
\begin{equation}
\label{nearly}
w_a(\mu) \propto Ai(r^{1/3} a (\mu+1/3))
\end{equation}
which solves our problem: from this in fact we see that, since $d^2 Ai(x)/d x^2 
= x Ai(x)$ by definition, and thus $d^2 Ai(x)/d x^2 = 0$ in $x=0$,  then we 
must have
\begin{equation}
\label{sospirata}
\frac{\partial^2 w_a}{\partial \mu^2} |_{\mu = -1/3} = 0\;.
\end{equation}
This is our sought for extra condition for $s$; we have seen that it comes 
directly from demanding that the boundary condition of the problem, Eq. 
\ref{incomplete}, manages to pass through the singular point of Eq. \ref{main},
which I showed to be a conventional one--turning point problem familiar from
elementary quantum mechanics. 

By substituting into Eq. \ref{incomplete} I find
\begin{equation}
\left(\frac{\partial^2 w_a}{\partial \mu^2}\right)_{\mu = -1/3} = 
2^{-2(2+s)} 3^{2+s} e^{-1} (s^2 -5s + 3) = 0
\end{equation}
\begin{figure}
\epsscale{1.0}
\plotone{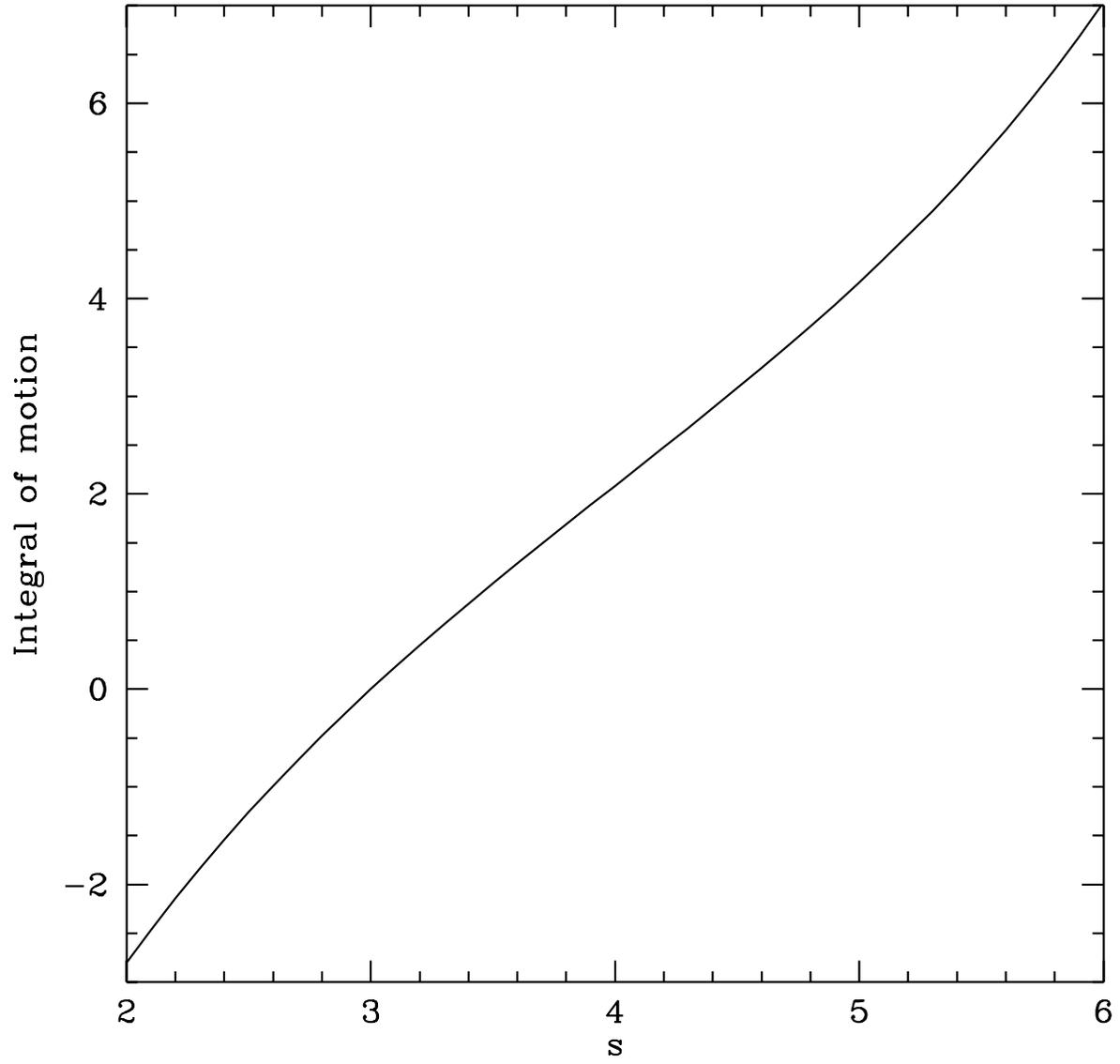}
\caption{The integral of Eq. \ref{conserved2}b with $u=1/3$, for the angular 
distribution of Eq. \ref{incomplete}, as a function of the parameter $s$, 
with arbitrary vertical scale.}
\end{figure}
from which we obtain $s = (5 \pm \sqrt{13})/2$. The solution with the 
minus sign is unacceptable: in fact, if we plug Eq. 
\ref{incomplete} into the conservation equation \ref{conserved1}, we see (Fig. 
1) that for $s \leq 3$ the integral is $\leq 0$. We remarked after Eq. 
\ref{conserved2}b, however, that this integral had necessarily to be $> 0$, so 
that we may conclude that $3$ is an absolute lower limit to $s$. Thus we discard
the solution with the minus sign, and are left with the unique solution
\begin{equation}
\label{ss}
s = \frac{5 + \sqrt{13}}{2} \approx 4.30\;.
\end{equation}

\subsection{Oblique shocks}

Let us call $\phi$ the angle that the magnetic field makes with the shock 
normal, in the upstream fluid. Then shocks can be classified as either
subluminal or superluminal, depending upon whether, upstream, $u/\cos\phi < 1$
or $u/\cos\phi > 1$, respectively (de Hoffmann and Teller 1950). We are 
interested in the limit $u\rightarrow 1$, so that most shocks will be of the
superluminal type. In this case, we could (but we won't) move to a frame where 
the magnetic field is parallel to the shock surface, both upstream and 
downstream. However, downstream this extremely orderly field configuration 
appears more idealized than warranted by physical 
reality and observations. In fact, behind a relativistic shock, a large number
of processes (compression, shearing, turbulent dynamo, Parker instability, 
two--stream instability) can generate magnetic fields; furthermore, there is
no obvious reason why these fields should have large coherence lengths. In 
GRBs, a large number of observations of different afterglows supports this 
picture, the most detailed of all being those of GRB 970508, extending from a 
few hours to $400\;d$ after the burst (Waxman, Frail and Kulkarni 1998; Frail, 
Waxman and Kulkarni 2000, and references therein). Accurate and successful
modeling fixes the 
post--shock ratio of magnetic to non--magnetic energy densities to $\epsilon_B 
\approx 0.1$. Notice that here the protons' rest mass is not even the largest
contribution to the non--magnetic energy density! Polarization measurements
also support, albeit less cogently, the idea of a small coherence length: of 
the four bursts observed so far, only one has a detected polarization, at the 
$1.7\%$ level (GRB 990510, Covino \etal, 1999). 

Thus the most plausible physical model downstream, is that particles move in a 
locally generated turbulent, dynamically negligible magnetic field; if then
we call $l$ the average post--shock field coherence length, and 
restrict our attention to particles with sufficiently large energies (\ie,
with gyroradii $r_g > l$), we see there can be no reflection as 
particles approach the shock from downstream. It follows that we expect the 
situation downstream to be identical to that of pure pitch angle scattering.
Upstream, the parallel magnetic field is also irrelevant. In fact, backward
deflection of a particle occurs on a length--scale $r_g$, but backward 
diffusion of the particle by magnetic irregularities only requires the
sideways deflection by an angle $\approx \gamma^{-2}$ ($\gamma$ being the shock
Lorentz factor), for the shock to overrun the particle. 
This typically occurs on a length $\eta r_g/\gamma^2$, 
with $\eta \approx $ a few. So, as $\gamma \rightarrow \infty$, the 
length--scale for scattering upstream by the magnetic field increases, while 
that by magnetic irregularities decreases: the field is irrelevant. In the end,
the same analysis as for pure pitch angle scattering applies, and the same
index $s$ and pitch angle distributions at the shock follow. 

In the case of subluminal shocks, a similar comment applies. Downstream, we 
expect on a physical basis the same situation as for superluminal shocks. 
Upstream, Eq. \ref{main} is replaced by (Kirk and Heavens 1989)
\begin{equation}
\gamma\cos\phi (u+\mu)\frac{\partial f}{\partial z} = \frac{\partial}{\partial
\mu} \left(D(\mu, p) (1-\mu^2)\frac{\partial f}{\partial \mu}\right) \;.
\end{equation}
to which the same analysis as in Subsection 2.1 can be applied. Thus we find
the same $s$ and pitch angle distributions at the shock as above. 

As a corollary, it may be noticed that the above argument also implies that
the results above are independent of the ratio $\kappa_\perp/\kappa_\parallel$,
the cross--field and parallel diffusion coefficients.

\section{Discussion}

For ultrarelativistic particles the energy spectral index $k$ is related 
to $s$ by 
\begin{equation}
\label{final}
k = s-2 = \frac{1 +\sqrt{13}}{2} \approx 2.30\;,
\end{equation}
\begin{figure}
\epsscale{1.0}
\plotone{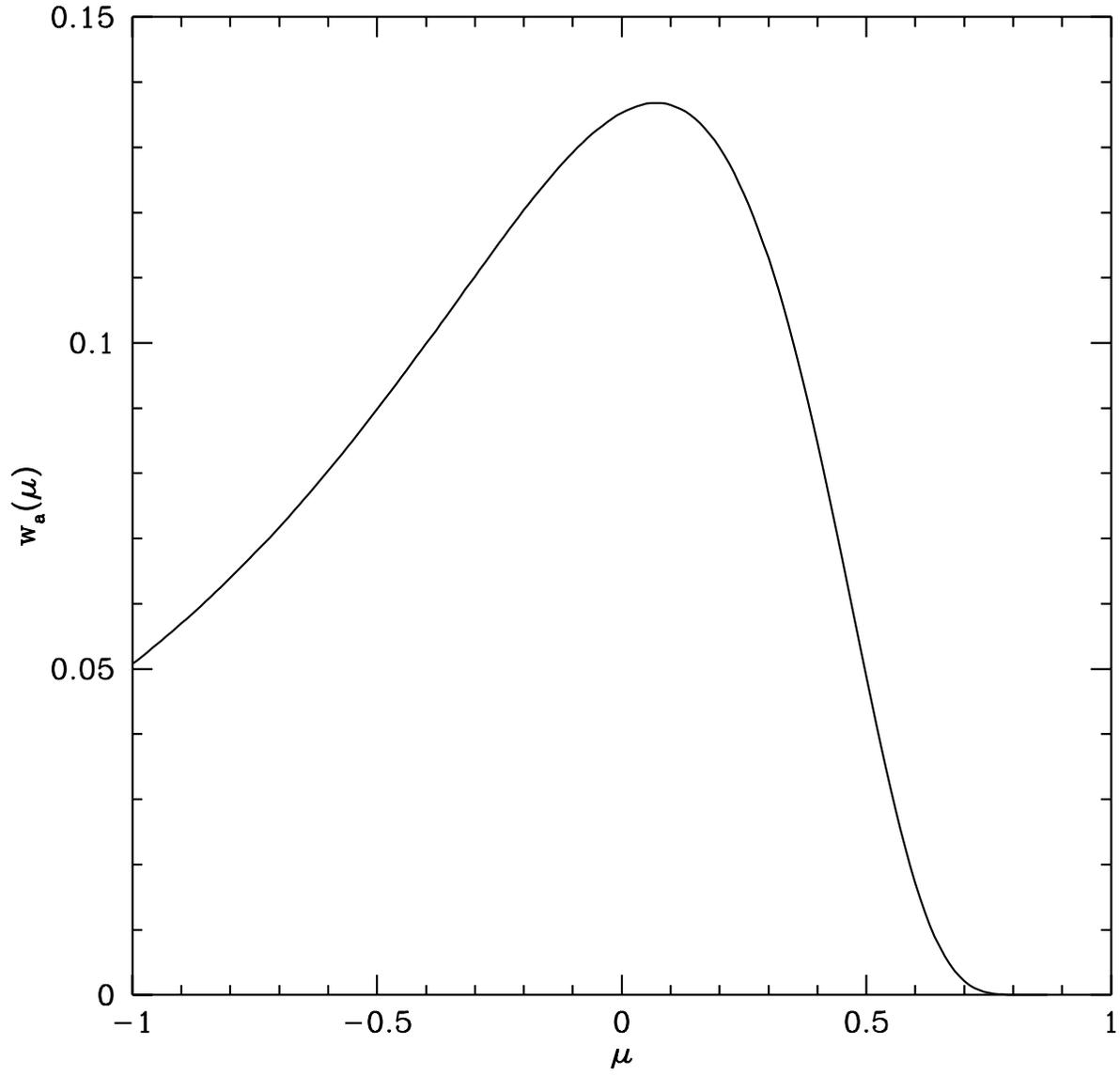}
\caption{Pitch angle distribution at the shock in the downstream frame, Eq. 
\ref{incomplete} with $s$ from Eq. \ref{ss}, with arbitrary vertical scale.}
\end{figure}
which is our final result. Also, now
that we know $s$, the final pitch angle distribution at the shock, but 
downstream, can be determined (Eq. \ref{incomplete}), and is plotted in Fig. 2.
None of these results depends upon the specific form of $D(\mu, p)$, so that
widely differing assumptions should yield precisely the same results. 

How does this compare with numerical work? The near--constancy of the index
$s$ (or $k$) explains moderately well the results of previous authors: 
Kirk and Schneider (1987) find $s= 4.3$ for their computation with the highest 
speed, which is however a modest Lorentz factor of $\gamma = 5$, and a single
functional dependence of $D$ on $\mu$. Heavens and Drury (1988) find again
a result of $s=4.2$ for equally modest Lorentz factors, but for 
two different recipes for $D$. Extensive numerical computations using a 
MonteCarlo technique (\ie, totally independent of the validity of Eq. 
\ref{main}) were performed by Bednarz and Ostrowski (1998), for a wide variety
of assumptions about the scattering properties of the fluids. They 
remarked quite explicitly that the energy spectral index $k$ seemed to 
converge to a constant value, independent of shock Lorentz factor (provided
$\gamma \gg 1$), magnetic field orientation angle $\phi$, and diffusion
coefficient ratio, $\kappa_\perp/\kappa_\parallel$. They found $k \approx 2.2$. 
The present 
work confirms (for all untried forms for $D(\mu, p)$) and extends their 
simulations (by yielding the exact value, and explicit forms for the particle 
angular distributions). The upstream angular distributions also agree well:
Fig.3a of Bednarz and Ostrowski clearly shows that this is (for the highest 
displayed value of $\gamma = 27$) a Dirac's delta, in agreement with the 
large--$\gamma$ limit in Section 2.1. However, the downstream pitch angle
distributions (in their Fig. 3b) agree well with mine, but not perfectly. 
Gallant, Achterberg and Kirk (1998) have claimed that there is a small error in 
Bednarz and Ostrowksi's distributions. As a matter of fact, my distribution
(Fig. 2) agrees much better with Gallant \etal's and Kirk and Schneider's
(1987) than Bednarz and Ostrowski's, despite the very small shock Lorentz
factors of these two papers ($\gamma = 2.3$ and $\gamma = 5$, respectively). 
Possibly, the small error in question may even explain the (small!) discrepancy 
between the two values of $k$.

A limitation applies to the claim of universality of Eqs. 
\ref{up}, \ref{incomplete} and \ref{final}: I
neglected any process altering the particles' energy during the scattering.
Clearly, the results of this paper only apply in the limit $\delta\!p/p \la 
1$, where $\delta\!p$ is the typical momentum transfer in each scattering 
event. In the large momentum limit considered here, it seems unlikely
that this constraint may be violated. 

Lastly, a comment on the assumed dependence $\propto p^{-s}$ of the 
distribution function upon particle momenta is in order. It can be seen from
Eq. \ref{main} that such a dependence is {\it not} required by this equation. 
To see this, let us make the usual assumption that $D$ is homogeneous of
degree $-r$ in $p$, \ie, $D(\mu, p) = q(\mu) p^{-r}$. Then by defining a new 
variable $\hat{z} \equiv z/p^r$, we see that the form assumed by Eq. \ref{main}
after this change of variable is identical to the original one, except that now
$p$ has altogether disappeared. At large $z$ (\ie, far downstream), $f 
\rightarrow f_\infty =$ constant, and there is no $p$--dependence. This 
paradox is solved by noticing that the real problem to be solved involves
both scattering (= Fermi acceleration) and injection. In this case, a
typical injection momentum $p_0$ arises naturally, and the dimensional problem
discussed above is naturally solved: we must have $f = f(...,p/p_0,...)$
where the dots indicate all other parameters. In the limit of $p_0 
\rightarrow 0$, $f$ {\it does not} tend to a constant independent of $p_0$
as is always assumed, but tends instead to zero as $f \rightarrow (p_0/p)^s$.
Problems of this sort, though rare in astrophysics, are common in hydrodynamics,
where they are called self--similar problems of the second kind (Zel'dovich
1956). They range from the deceptively simple laminar flow of an ideal fluid
plast an infinite wedge (Landau and Lifshitz 1987) to the illuminating case of
the filtration in an elasto--plastic porous medium (Barenblatt 1996). It is
remarkable that, in the problem at hand, no such complication is necessary to
fix the all--important index $s$, yet the powerful methods of intermediate
asymptotics (Barenblatt 1996) and the renormalization group (Goldenfeld 1992)
can be brought to bear on the intermediate $\gamma$ cases, where no easy 
limiting solution can be found. 

In short, what I have done in this paper is to show that the spectrum of 
non--thermal particles accelerated at relativistic shocks is universal, in
the sense that the energy spectral index $k$, and the angular distributions
in both the upstream and downstream frames (Eqs. \ref{up}, \ref{incomplete},
\ref{final}, and Fig. 2) do not depend upon the scattering function $D(\mu, p)$,
the shock Lorentz factor (provided of course $\gamma \gg 1$), the magnetic
field geometry, and the ratio of cross--field to parallel diffusion 
coefficients.Thus we have the result that the cosmic rays' spectra are 
independent of flow details in both the Newtonian (Bell 1978) and the 
relativistic regimes. 

{}

\end{document}